\documentclass[twocolumn]{article} % default is 10 pt
\usepackage{graphicx} % needed for including graphics e.g. EPS, PS

\long\def\comment#1{}

\makeatletter
\def\@normalsize{\@setsize\normalsize{10pt}\xpt\@xpt
\abovedisplayskip 10pt plus2pt minus5pt\belowdisplayskip
\abovedisplayskip \abovedisplayshortskip \z@
plus3pt\belowdisplayshortskip 6pt plus3pt
minus3pt\let\@listi\@listI}

\def\subsize{\@setsize\subsize{12pt}\xipt\@xipt}
\def\section{\@startsection {section}{1}{\z@}{1.0ex plus
1ex minus .2ex}{.2ex plus .2ex}{\large\bf}}
\def\subsection{\@startsection
   {subsection}{2}{\z@}{.2ex plus 1ex} {.2ex plus .2ex}{\subsize\bf}}
\makeatother

\newtheorem{fig}{\bf Figure}

\begin{document}

% don't want date printed
\date{}

% >>>>>>>>>>>>>>>>>>>>>>>  Put your title here <<<<<<<<<<<<<<<<<<<<<<<<
% make title bold and 14 pt font (Latex default is non-bold, 16pt)
\title{\huge \bf {Navigation in tilings of the hyperbolic plane
and possible applications}}

% >>>>>>>>>>>>>>>>>>>>>>> Author's Name, Thanks or Affliation <<<<<<<<
\author{Maurice Margenstern
 \thanks{
%Text in here gets placed in a footnote section.  Typically you
% thank sponsors here, and add the date of the manuscript 
%submission and your address e.g.
% Drexel University Mechanical Engineering, Robotics
% \& Machine Vision Lab, Philadelphia PA USA 19104
% Tel/Fax: 215-895-6396/1478 Email: paul@coe.drexel.edu
Universit\'e Paul-Verlaine $-$ Metz, \vskip 0pt
LITA, EA 3097, UFR MIM,\vskip 0pt
Campus du Saulcy, 57045 METZ C\'edex, FRANCE, 
\vskip 0pt
Email: margens@univ-metz.fr
 }
}

\maketitle
\thispagestyle{empty}

%\subsection*{}

% >>>>>>>>>>>>>>>>>>>>>>>>> Keywords and Abstract <<<<<<<<<<<<<<<<<<<<<
% Replace with your own keywords and abstract.  Text will be in italics
{\hspace{1pc} {\it{\small Abstract}}{\bf{\small---This paper
introduces a method of navigation in a large family of tilings
of the hyperbolic plane and looks at the question of possible
applications in the light of the few ones which were already 
obtained.
%ese instructions
%give you guidelines for preparing papers for the IMECS. Use this
%document as a template if you are using LaTeX. Motion tracking and
%object recognition often use cameras that are mounted in motion
%platforms like pan-tilt units, linear tables and even robots.
%Tracking can be automated by visually servoing the platform's
%degrees-of-freedom (DOF) thus keeping the camera's point-of-view
%directed at the target. Tracking quick moving targets often demands
%faster bandwidth platforms. However biology suggests a redundant
%approach where DOF, like the eye and head, cooperate to direct
%vision systems and overcome joint limits.  This paper illustrates
%the effectiveness of this concept using a robot-mounted camera.

\em Keywords: hyperbolic tilings, cellular automata,
applications.
%visual-servoing, tracking, biomimetic, redundancy,
%degrees-of-freedom 
}}
}

% >>>>>>>>>>>>>>>>>>>>>> START OF YOUR PAPER <<<<<<<<<<<<<<<<<<<<<<<<<<<<<<
% Typically paper starts of with an Introduction header.  
%Replace text in
% the french braces if you see fit.  I also typically name the 
%label the
% same as the section

\section{Introduction}
\label{Introduction}

   Hyperbolic geometry appeared in the first half of the 
19$^{\rm th}$ century, proving the independence of the parallel
axiom of Euclidean geometry. Models were devised in the second
half of the 19$^{\rm th}$ century and we shall use here one of the
most popular ones, Poincar\'e's disc. This model is represented by 
Figure~\ref{poincare_disc}.
\vskip 7pt
\vtop{
%\begin{figure}[b]
\centerline{
\mbox{\includegraphics[width=200pt]{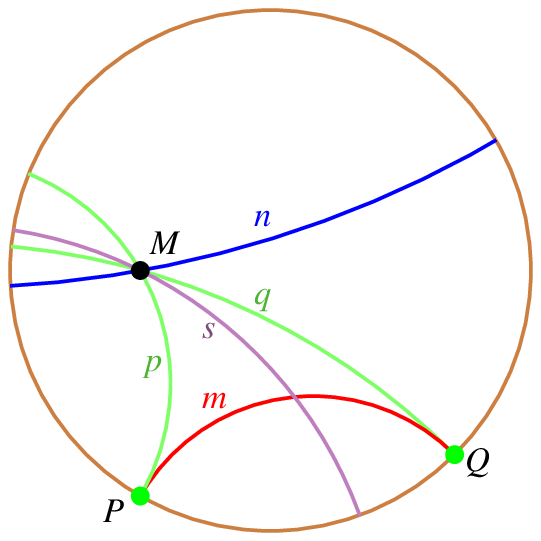}}
}
\vspace{-20pt}
\begin{fig}
\label{poincare_disc}\small
Poincar\'e's disc model: inside the open disc, the points of the hyperbolic plane.
Lines are trace of diameters or circles orthogonal to the border of the disc, e.g. 
the line~$m$. Through the point~$A$ we can see a line~$s$ which cuts~$m$, two lines
which are parallel to~$m$: $p$ and~$q$, touching~$m$ in the model at~$P$ and~$Q$ 
respectively which are points of the border and which are called points at infinity.
At last, and not the least: the line~$n$ also passes through~$A$ without cutting~$m$,
neither inside the disc nor outside it.
\end{fig}
}
\vskip 7pt
   From a famous theorem established by Poincar\'e in the late 19$^{\rm th}$ century, 
it is known that there are infinitely many tilings in the hyperbolic plane, each one
generated by the reflection of a polygon~$P$ in its sides and, recursively, in the
reflection of the images in their sides, provided that the number~$p$ of sides of~$P$
and the number~$q$ of copies of~$P$ which can be put around a point~$A$ and exactly
covering a neighbourhood of~$A$ without overlapping satisfy the relation:
\hbox{$\displaystyle{1\over p}+\displaystyle{1\over q}<\displaystyle{1\over2}$}.
The numbers $p$ and~$q$ characterize the tiling which is denoted $\{p,q\}$ and
the condition says that the considered polygons live in the hyperbolic plane. Note
that the three tilings of the Euclidean plane which can be defined up to similarities
can be characterized by the relation obtained by replacing~$<$ with~$=$ in the above
expression. We get, in this way, $\{4,4\}$ for the square, $\{3,6\}$ for the equilateral
triangle and $\{6,3\}$ for the regular hexagon.  

   In the paper, we shall focus our attention on the simplest tilings which can be
defined in this way in the hyperbolic plane: $\{5,4\}$ and~$\{7,3\}$. We call them 
the {\bf pentagrid} and the {\bf heptagrid} respectively, see Figure~\ref{tilings_54_73}.
\vskip 7pt
\vtop{
\centerline{
\mbox{\includegraphics[width=120pt]{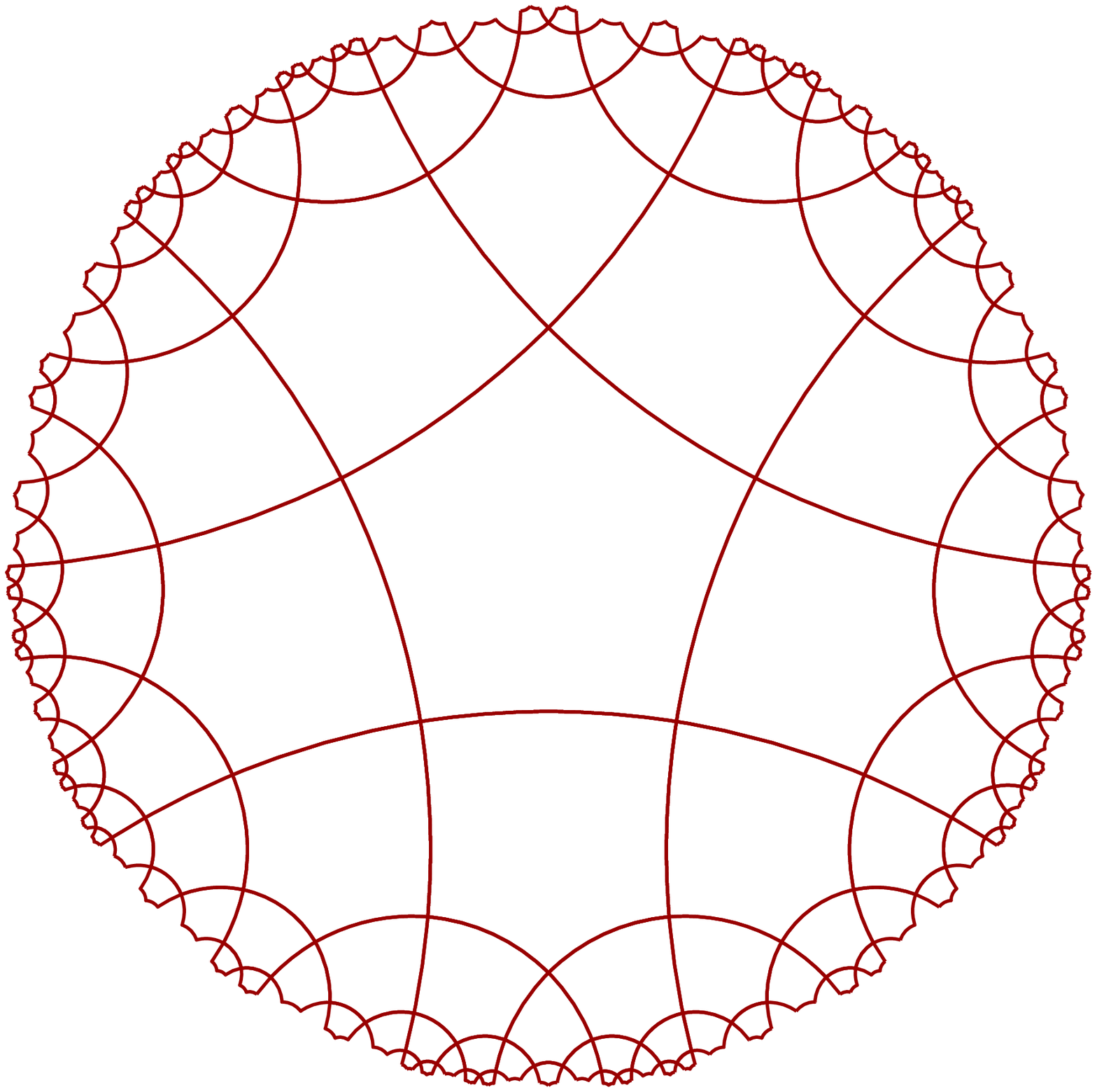}}
\mbox{\includegraphics[width=120pt]{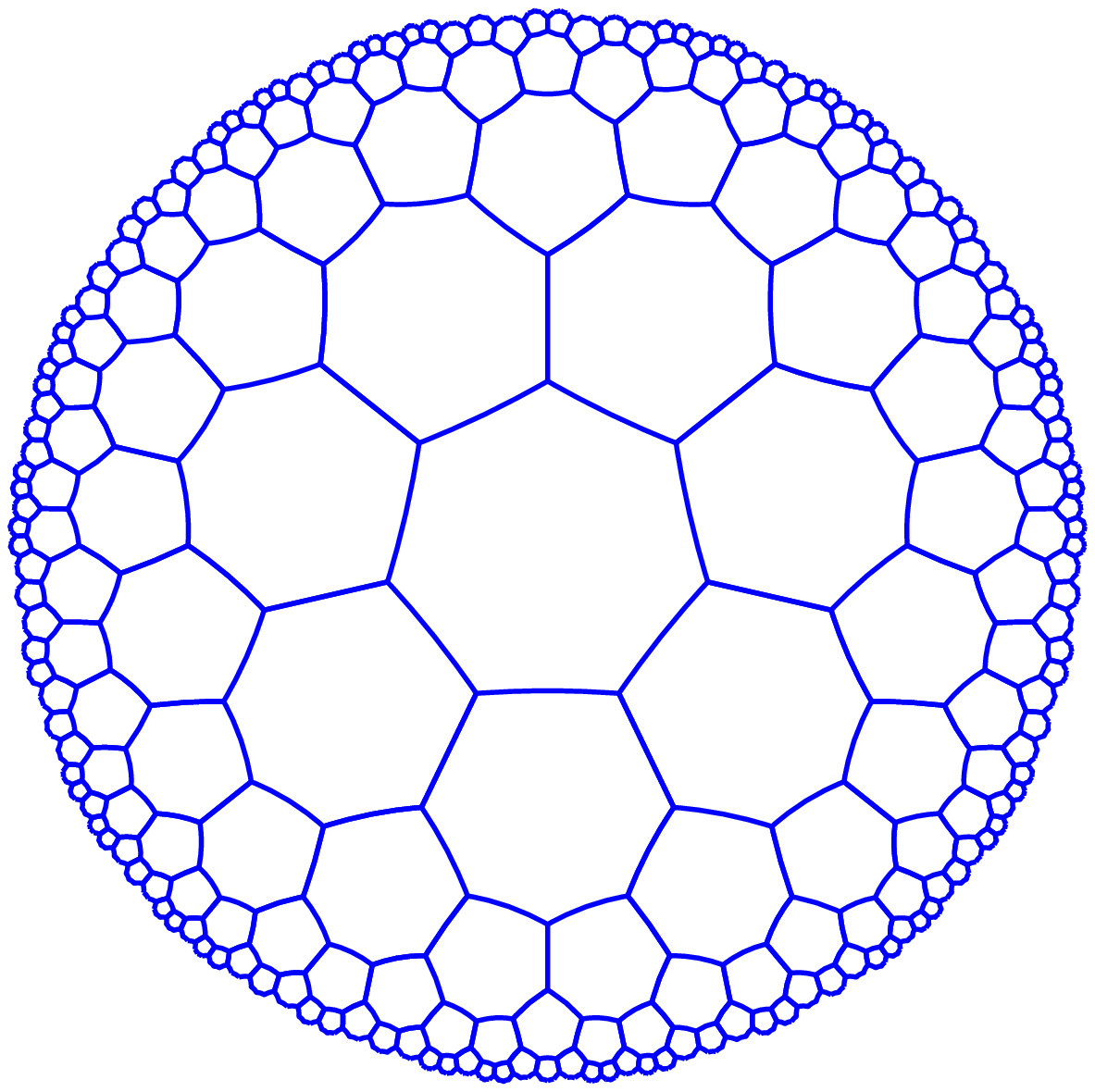}}
}
\vspace{-10pt}
\begin{fig}
\label{tilings_54_73}\small
Left-hand side: the pentagrid; right-hand side: the heptagrid.
\end{fig}
}
%\vskip 7pt

   To navigate in this tilings was for a long time a non trivial question. In 1999 and
2000, see~\cite{mmkm99,mmJUCSii}, the author found a technique which allows to find 
one's way in these tilings, first in the pentagrid. The generalization of this way to 
the heptagrid and to infinitely other tilings of the hyperbolic plane was obtained a bit 
later. Details and references on these results can be found in~\cite{mmbook1,mmbook2} 
as well as their extension to one tiling of the hyperbolic $3D$~space and another 
one of the hyperbolic $4D$~space.
  
   Cellular automata are a tool used in
various sciences, from gas statistical physics to economy, for simulation purposes with
good results and a few industrial applications. We refer the reader to proceedings
of the last two issues of {\bf ACRI} conferences to have a look at this range of
applications.
   
   The navigation technique introduced in~\cite{mmJUCSii} allowed to implement cellular
automata in the pentagrid and in the heptagrid and to devise a few applications. 

   In Section~\ref{Navigation}, we sketchily describe the navigation technique.
In Section~\ref{CellularAutomata}, we remind the results on cellular automata and in
Section~\ref{Applications}, we consider the applications already performed as well as
a few others which could be useful. In Section~\ref{Conclusion}, we conclude with
what could be done in future work.

%\caption{
%\end{figure}
%}
%\mbox{\includegraphics[width=2.00in]{engineeringletters.eps}} }
%\caption{Replace text here with your desired caption.}
%\label{overView}
%\end{figure*}

\section{Navigation in the pentagrid, in the heptagrid} %and in the dodecagrid}
\label{Navigation}

   The navigation in a tiling of the hyperbolic space can be compared to the flight
of a plane with instruments only. Indeed, we are in the same situation as a pilot
in this image as long as the representation of the hyperbolic plane in the Euclidean one
entails such a distortion that only a very limited part of the hyperbolic plane
is actually visible.

\vskip 7pt
\vskip 7pt
\vtop{
\centerline{
\mbox{\includegraphics[width=180pt]{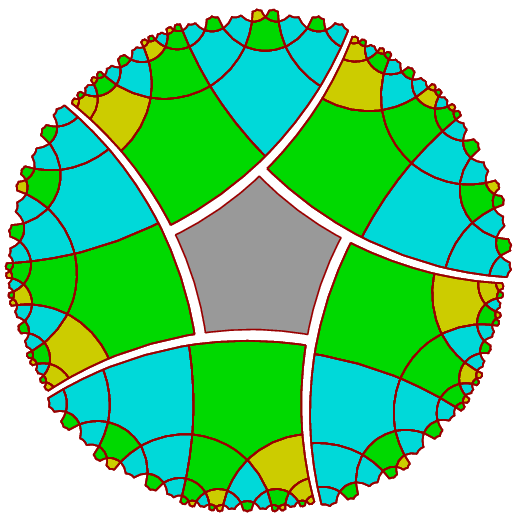}}
}
\vspace{-15pt}
\begin{fig}
\label{eclate_54}\small
First part of the splitting: around a central tile, fixed in advance, five sectors.
Each of them is spanned by the Fibonacci tree defined in Figure~{\rm\ref{split_54}}.
\end{fig}
}
%\vskip 7pt

   The principle of the navigation algorithms rely on two ideas which we illustrate
on the pentagrid. The first one, is a way to split the hyperbolic plane, see 
Figure~\ref{eclate_54} and Figure~\ref{split_54}: the recursive structure of this
splitting defines a tree which spans the tiling. The second idea consists in numbering
the nodes of the tree level after level and, remarking that the number of nodes on the
level~$n$ is $f_{2n+1}$ defined by the Fibonacci sequence where $f_0=f_1=1$, to represent
the numbers in the numbering basis defined by this sequence. Also, as the representation
is not unique, we fix it by choosing the longest one. 

   On Figure~\ref{split_54}, we can notice that the Fibonacci tree has two kinds of nodes:
the {\bf white} ones, which have three sons, and the black ones, which have two sons.
In both cases, the leftmost son is black, the others are white. Figure~\ref{fibo}
represents the tree in a more traditional way together with the numbering of the nodes
and their representation in the Fibonacci basis. Later on, we shall call {\bf coordinate}
this representation of the number of a node.
\vskip 7pt
\vtop{
\centerline{
\mbox{\includegraphics[width=120pt]{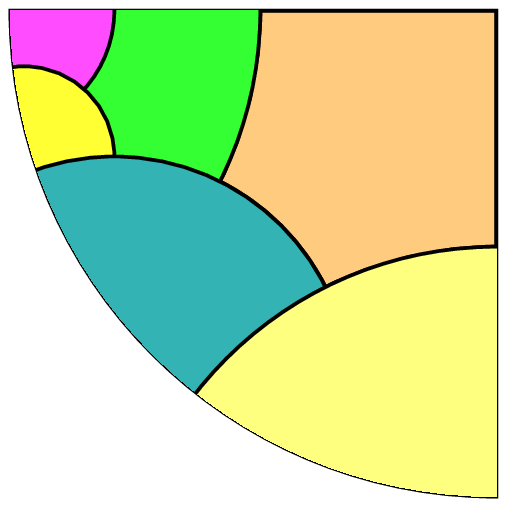}}
\mbox{\includegraphics[width=120pt]{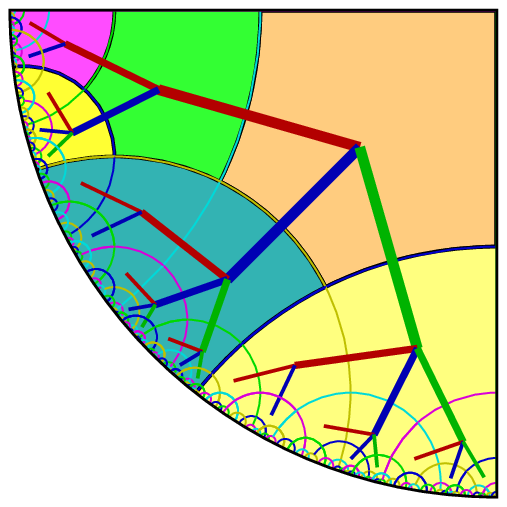}}
}
\vspace{-20pt}
\begin{fig}
\label{split_54}\small
Second part of the splitting: splitting a sector, here a quarter of the hyperbolic plane.
On the left-hand side: the first two steps of the splitting. On the right-hand side: 
expliciting the tree which spans a sector.
\end{fig}
}
%\vskip 7pt

   There are important properties which are illustrated by Figure~\ref{fibo} 
and which I called
the {\bf preferred son properties}. It consists in the fact that for each node of a Fibonacci
tree, among the coordinates of its sons, there is exactly one of them which is obtained 
from the coordinate of the node by appending two 0's and which is called the preferred
son. Moreover, the place of the preferred son is always the same among the sons of a node:
the leftmost son for a black node, the second one for a white node.
\vskip 7pt
\vtop{
\centerline{
\mbox{\includegraphics[width=210pt]{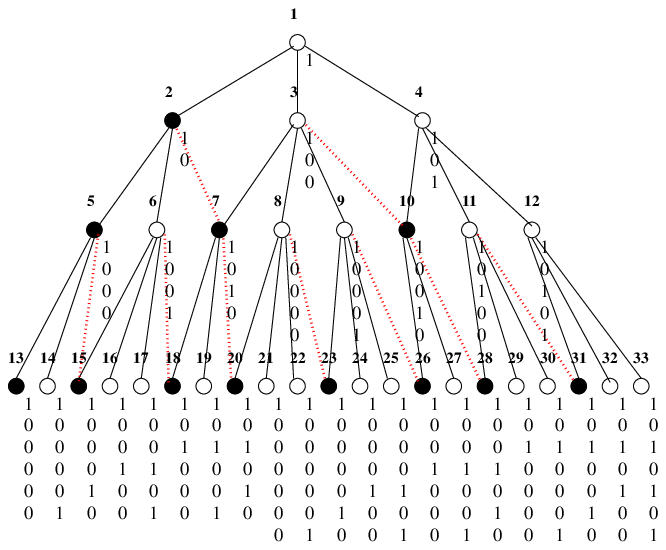}}
}
\vspace{-20pt}
\begin{fig}
\label{fibo}\small
The Fibonacci tree: the representation of the numbers of the nodes in the Fibonacci
basis. 
\end{fig}
}
%\vskip 7pt

   From the preferred son properties, it was possible to devise an algorithm which
computes the path from the root to a node in a linear time in the length of the 
coordinate of the node, see \cite{mmASTC}. From this, we also get that the coordinates
of the neighbours of a given tile can be computed from the coordinate of the tile in
linear time too.

\vskip 7pt
\vskip 7pt
\vtop{
\centerline{
\mbox{\includegraphics[width=180pt]{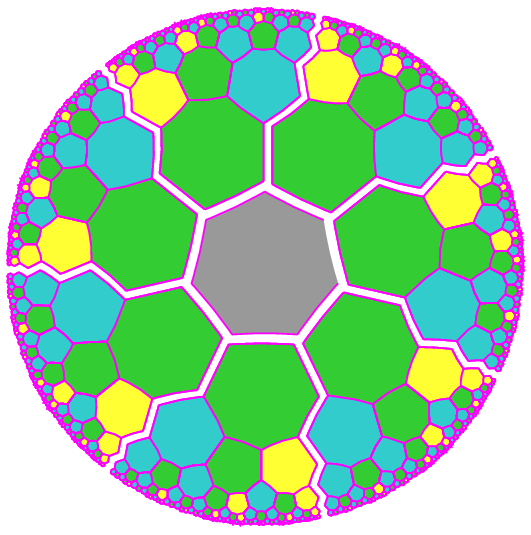}}
}
\vspace{-15pt}
\begin{fig}
\label{eclate_73}\small
First part of the splitting: around a central tile, fixed in advance, seven sectors.
Each of them is spanned by the Fibonacci tree defined in Figure~{\rm\ref{split_54}},
see Figure~{\rm\ref{mid-point}}.
\end{fig}
}
%\vskip 7pt

   It is interesting to remark that the heptagrid possesses properties which are very 
similar to those of the pentagrid, see Figures~\ref{eclate_73}
and~\ref{mid-point}. Note that Figure~\ref{mid-point} explains why the same tree basically 
spans each of the seven sectors of Figure~\ref{eclate_73}. 
\vskip 7pt
\vtop{
\centerline{
\mbox{\includegraphics[width=180pt]{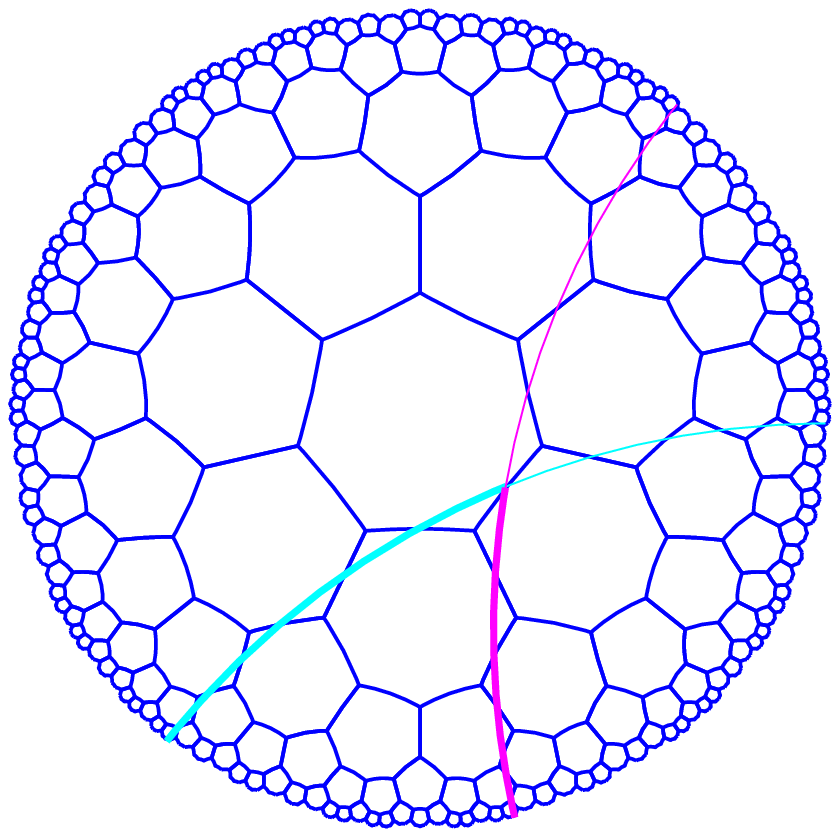}}
}
\vspace{-15pt}
\begin{fig}
\label{mid-point}\small
The mid-point lines: the tool which shows how a sector spanned by the Fibonacci tree is 
defined in the heptagrid.
\end{fig}
}
%\vskip 7pt

   We conclude this section by mentioning that these nice properties of the pentagrid and
of the heptagrid can be extended to two infinite families of tilings of the hyperbolic 
plane. In particular, the fact that the same tree spans the pentagrid and the heptagrid
can be extended as follows: for each $p$, $p\geq5$, the same tree spans the 
tiling $\{p,4\}$ and the tiling $\{p$+$2,3\}$. The trees are different for different
values of~$p$ although they share a common feature: nodes are divided into black and
white nodes. The difference in the number of sons from a white node to a black one
is~1. The unique black son can be chosen to be the leftmost one.

\section{Results on cellular automata in hyperbolic spaces}
\label{CellularAutomata}

   These fast algorithms allowed to implement cellular automata in the hyperbolic plane,
first in the pentagrid and, a few years later, in the heptagrid. 

   We have three kinds of results regarding cellular automata in these spaces: complexity 
results, universality results and a solution of two problems dealing with communications
between cells of a cellular automaton. Although these results have a definite theoretical
character, they nevertheless have a practical significance. 

   The most striking result regarding complexity, is that for cellular automata in the
hyperbolic plane, we have \hbox{\bf P$_{hc}$ = NP$_{hc}$}. This means that non-deterministic 
polynomial time computations of a cellular automaton in the hyperbolic plane can be 
performed also in polynomial time by a cellular automaton in the hyperbolic plane.
Moreover, the exact power of computation of the class {\bf P}$_{hc}$ is the well-known 
class {\bf PSPACE}. The reader is referred to~\cite{mmkm99,immwSCI,cimmIEICE,mmbook2}.

   For what is universality, there are universal cellular automata in the hyperbolic
plane, {\it i.e.} able to simulate the computation of any Turing machine. What is
more important is that if infinite but elementary initial configurations are allowed
$-$ within this limited room we cannot formally describe the exact meaning of this
expression $-$ it is possible to simulate any Turing machine with a cellular automaton
with 9 states in the case of the pentagrid, see~\cite{mmjsPPL}, and with 6 states
in the case of the heptagrid, see~\cite{mmjsENTCS} and even less: 4 states, 
see~\cite{mmRodos}. All the quoted cellular automata have a common structure:
they simulate a railway circuit traversed by a unique locomotive consisting of tracks
and three kinds of switches devised in~\cite{stewart}, see also~\cite{mmbook2}.

   Now, the linear algorithm for finding the path from a node to the root of the sector
to which the node belongs allows to devise efficient communication protocols for cells
of a cellular automaton in the pentagrid or in the heptagrid, 
see~\cite{mmJCA,mmJCAb,mmbook2}. 

   If the latter point is closer to applications, the first ones also say something on this
regard. The meaning of the complexity results is that hyperbolic cellular automata may run
much faster than their Euclidean analogues, as they have at their disposal an exponential
area which can be constructed and used in linear time. Also, these computations may be
universal as this is the case for their Euclidean analogue~2. And so, we can do 
any computations in this frame, never in more time than what is required for a Euclidean 
cellular automaton, and very often in much less time. In particular, in this setting,
hard problems of everyday life turn out to be solvable in polynomial time, very often 
even in linear time. 

\section{On the side of applications}
\label{Applications}

   The above results might seem to beautiful to be true. However, they are theoretical
results whose proofs were checked, some of them with the help of a computer program,
and they are correct.

   But is this feasible?

   Our local environment is usually thought as Euclidean although it would be better
to see it as governed by spherical geometry. Many cosmologists consider our universe
as a space with a negative curvature. In this regard, hyperbolic geometry could be a
better model for medium scale than Euclidean geometry. Computations on the orbit of
Mercury are more conformal to observations when performed in a hyperbolic setting.
So, our tools might have applications at this scale which is not a today urgent matter,
but we know that possible tools exist.

   I would also mention another argument. An important feature of hyperbolic geometry
is the lack of similarity. As a consequence, it can be said that a shape has necessarily
a certain size. As an example, in the hyperbolic plane, there is a unique size of the 
edge for a regular pentagon with right angles. This means that two such pentagons can
be transformed from each other through a simple geometric transformation which is an
isometry. And so, up to isometries, such a pentagon is unique. This property is shared
by any figure of the hyperbolic plane. Now, if we look at biology, we can notice that there
is no true similarity. Individual differences, for instance, are not answerable by 
similarity. This remark led me to~\cite{mmHypBio} where I used the pentagrid to
represent a theoretical model of living cells, the common point being the fact that, 
in both cases, a tree underlies the structures.

   Now, another field of application, more important in my eye, is provided by computer 
science itself. We need new concepts to handle problems raised by massive computations 
and by the management of huge amounts of information. For this, we need new horizons 
and it is not at all unreasonable to consider tilings of the hyperbolic plane with 
their navigation tools as a possible model for tackling these problems. Note, for 
instance, that trees are already used in the organization of operating systems: many a
user is faced with the tree structure of the directories of his/her machine. Now, 
trees are also spanning hyperbolic geometry and this is particularly blatant in the 
field of tilings in the hyperbolic plane as I hope the reader is convinced after reading 
Section~\ref{Navigation}.

   In the present section, we first have a look at the existing applications, see
Subsections~\ref{Colour} and \ref{Keyboards}. These applications are based on a
{\bf fisher eye} effect of the pentagrid and the heptagrid. Now, it is known that
fisher eye techniques are of interest for several applications in human-machine 
interaction. In Subsection~\ref{Others}, we again address the problem of representing,
storing and exploring information which mainly use the coordinate system.

\subsection{The colour chooser}
\label{Colour}
   The colour chooser is a software which was developed in my laboratory by members of
my team, see~\cite{ibkmIHM,mmbook2}. From the initial state of the chooser illustrated by
Figure~\ref{palette}, the user can navigate in the tiling in order to look after the
colour he/she thinks the most adapted for his/her purpose.

\vskip 7pt
\vtop{
\centerline{
\mbox{\includegraphics[width=210pt]{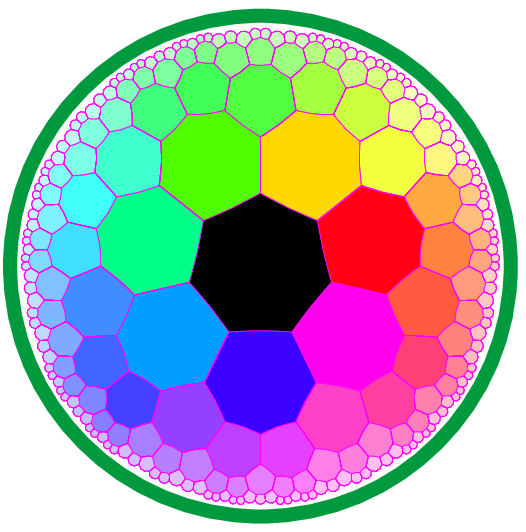}}
}
%\vspace{-15pt}
\begin{fig}
\label{palette}\small
The colour chooser: the pentagrid is also a possible tool but the heptagrid gives the 
best representation. Probably because at first glance, the heptagons of the figure are 
seen as hexagons: it is needed to count the sides in order to detect the difference.
\end{fig}
}
%\vskip 7pt

   There are seven fixed in advance keys which indicate to which tile the user wishes
to go from the central tile. Once the appropriate key is pressed, the chosen tile comes
to the centre of the disc and all the other tiles move correspondingly. Consequently,
the navigation appears as if the green disc of Figure~\ref{palette} would be a window 
moving over the hyperbolic plane. Once the black tile indicating the initial center is
no more visible, it is very easy to get lost. To avoid such a defect, the chooser keeps
an arrow pointing at a point of the green disc to which the user have to go in order
to go back to the initial centre.

\subsection{The keyboards}
\label{Keyboards}
   Another application was developed in the laboratory, see~\cite{bmKeyboard}, which
we consider in Subsection~\ref{Latin} and which was developed in a different direction,
as will be seen in Subsection~\ref{Japanese}.
 
\subsubsection{Latin Keyboard}
\label{Latin}

   The first idea was a proposal of a keyboard for cell phones. As our laboratory is
in France, tests were performed with French students on randomly chosen sequences
of French sentences devised for the experiment.

   Below, Figure~\ref{latinkeyboard} illustrates the basic principle of the software.
The idea is close to that of the chooser. But this time, the tiles contain letters
and the user goes from the initial empty centre to the desired letter by pressing
appropriate keys. At most three keys have to be pressed and with other media, pressing
the keys can be replaced by three quick moves of the hand. 

\vskip 7pt
\vtop{
\centerline{
\mbox{\includegraphics[width=80pt]{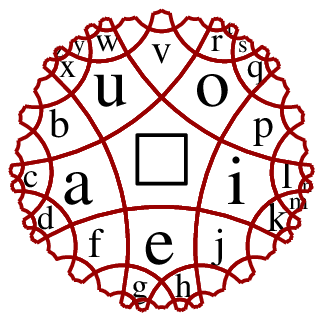}}
\mbox{\includegraphics[width=80pt]{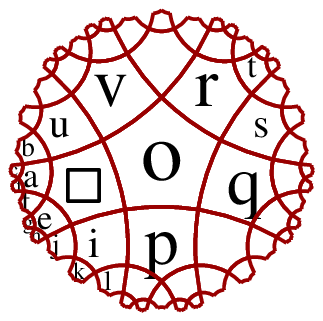}}
\mbox{\includegraphics[width=80pt]{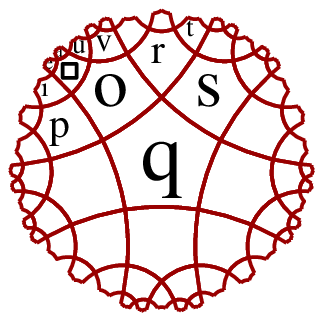}}
}
\vspace{-15pt}
\begin{fig}
\label{latinkeyboard}\small
A proposal of a keyboard for cell phones. 
\end{fig}
}
%\vskip 7pt

   Also, the pentagrid is used instead of the heptagrid because for this purpose it is 
more suited to the gestures of the user and letters can be better seen. The reason is that
the regular pentagon with right angle is bigger than the rectangular heptagon with
the angle $\displaystyle{{2\pi}\over3}$. This explains that letter can be better seen.
This explains also that less accuracy is required from the user for his/her gestures.
The experiments proved that the keyboard is certainly not worse than commercial products.
Moreover, among young people, it raised a curiosity which contributed to the quick learning
of the keyboard. Another important feature is that we adopted a distribution of the 
letters as close as possible to the standard alphabetic order, which raises no additional
effort of memory from the user.

\subsubsection{Japanese Keyboard}
\label{Japanese}
    The use of the pentagrid and the fact that hiraganas and katakanas of the Japanese
language are traditionally presented in series based on the five vowels of the language
inspired me the idea to use the same grid but this time, in a Japanese environment.
We did some work in this direction, with Japanese colleagues, see~\cite{mbuynACRI}. 
Figure~\ref{japanesekeyboard} illustrates the principle of the working.

   This keyboard has always met a big success in the conferences were I presented it in 
Japan. There was a prototypical implementation on actual cell phones and concluding
experiments were performed with a small group of Japanese students on a protocol
which was similar to the one followed in France but, of course, adapted to the
Japanese language.

   It is important to indicate that our project also aims at a full representation of the
kanjis used in the Japanese language, starting from a phonetic approach. This is
a very complex task and, at the present moment it is not completed. 
   
\vskip 7pt
\vtop{
\centerline{
\mbox{\includegraphics[width=210pt]{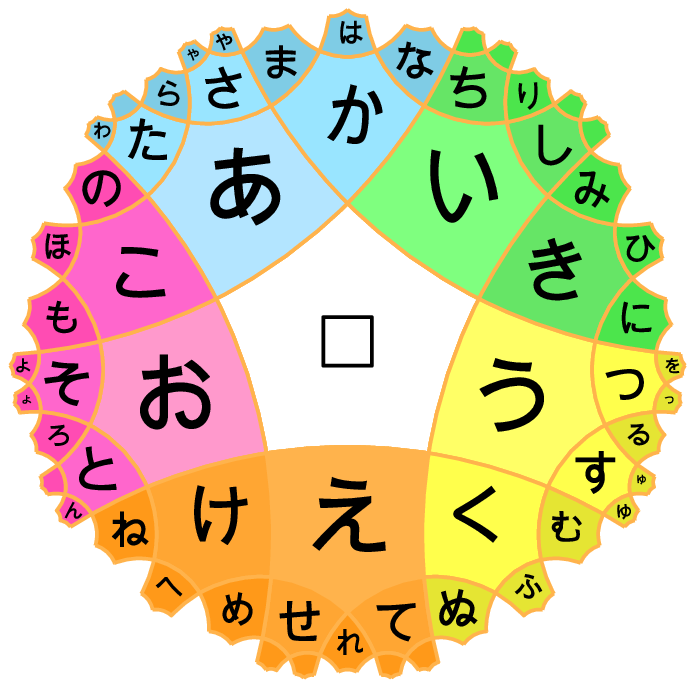}}
}
\vspace{-15pt}
\begin{fig}
\label{japanesekeyboard}\small
A proposal of a keyboard for Japanese cell phones. 
\end{fig}
}
%\vskip 7pt

   The question could be raised of the adaptation of the same principle to other languages.
I was indicated by several colleagues that the occurrence of exactly five vowels is
a common feature of many Asian languages as the languages from Malaysia and Philippines,
also including the Polynesian languages. It would be interesting to see whether similar
ideas could be developed for other languages: any proposal would be welcomed!
 
\subsection{Other possible applications}
\label{Others}

   Now, let us turn to other possible applications which were not yet tested and
which are based on other principles.

   As indicated in the beginning of this section, computer science and computer 
engineering could be an important field of application of the navigation technique 
introduced by Section~\ref{Navigation}.

   There is not enough room to develop such applications in a detailed way. This is 
why I shall mention three possible sub-fields of application and give general arguments
only in favour of these proposals.

   These three sub-fields are: the representation of the Internet, computer architecture  
and operating systems and data processing. 

   The Internet can be represented as a graph. This is often performed in force oriented
representation which assigns a mass to each node and define their respective positions
by application of the laws of mechanics. Using the pentagrid or the heptagrid can be an 
alternative representation. In~\cite{mmbook2}, I illustrate this point by defining 
addresses of nodes which is based on the navigation technique of Section~\ref{Navigation}
but which looks like the today used IP numbers. There is another advantage in this proposal.
For both the pentagrid and the heptagrid, \cite{mmbook2} gives an algorithm which,
from the coordinates of two tiles~$A$ and~$B$, gives a shortest path from~$A$ to~$B$ in 
linear time in the size of the coordinates. This is an important improvement of the
path algorithm from a node to the root of the sector of the node mentioned in 
Section~\ref{Navigation}. It also allows to define a fast algorithm for changing
the centre of coordinates.

   This possibility may better represent the Internet connections by giving addresses
which are continuous with respect to the connection distance between nodes. 
In~\cite{mmJCAb}, I introduced a protocol of communication between tiles of the
pentagrid or the heptagrid based on the following principle: there is an absolute
system of coordinates but when sending a message, a tile considers itself as the central
tile and sends a message to all tiles together with its {\bf relative} address which is~0. 
Now each tile which receives the message, convey it to its sons with respect to the 
coordinate system of the sender, appending an information of constant size in order to
form the address it will convey to its sons. This allows a tile which receives the message
and which wishes to establish a contact with the sender to send an answer to the 
appropriate sender: it simply reverses the address in a way which is described 
in~\cite{mmJCAb}. 

    This can be transported to a set of processors: they could
be organized as if they would stand in a disc of the pentagrid or of the heptagrid around
a central one which would not necessarily have a control function. It is enough for that
to assign them coordinates as those introduced in Section~\ref{Navigation}. The 
communication between processors could be organized according to the just described
protocol.

   The shortest path algorithm can be used for the representation of the 
Internet but also for the representation of a file storage for an operating system. 
The system of coordinates which allow to construct a shortest path from one point to 
another in a linear time with respect to both coordinates should facilitate the finding 
of queries as topological neighbourhood is up to a point reflected in the coordinates 
themselves. From this remark, we can see that this can be of help also for data 
processing. In particular, constant saving mechanisms can be organised by scanning a 
circular area by a branch which moves from a central point to an indicated point of 
the circumference. The idea is that the branch constantly moves around this circumference 
and that at each time, the content of the branch only is saved. After a certain time, 
everything is saved and updates can also be included in this constant saving provided that 
it does not exceed a fixed in advance amount. In case of exceeding the threshold, the 
update is split in as many parts as needed in order that each part should fall within the 
threshold. 
%Within the threshold, it can also be decided to give priority to the update if it 
%concerns a point of the branch which is saved by the constant process while a more 
%recent update occurred.
It can be noted that when the branch goes from one node of the circumference to the next 
one, in many cases, the two positions of the branch may have a long common interval~$I$ 
starting from the center which consists of the same nodes. If no update occur on the
nodes belonging to~$I$ between the two corresponding tops of saving, then only a small
part of the new branch has to be saved. This can be easily managed by an appropriate 
signalization on the branch, see~\cite{mmJCAc}.

%  communication again with the result on the shortest path
\section{Conclusion}
\label{Conclusion}

   There is still much work to do in this domain, both in theory but also for applications.

   Practical problems are difficult as this is witnessed by the example of scheduling 
problems, either in airplane traffic or in production processes: such problems are
NP-complete. We have seen that the frame proposed by this paper allows to solve them in
polynomial time. This is not the single theoretical approach leading to such results.
As an example, molecular computations based on a modelling of DNA strand reactions
or on a modelling of a living cell lead to similar results. But our frame has the 
advantage of not being concerned by still unsolved biological problems for a practical 
implementation, the nature of which is not yet well understood: either it comes from
fundamental issues or it comes from not well enough mastered techniques.
Subsection~\ref{Others} pointed at fields where our approach might have feasible 
applications.

   It seems to me that exchanging all possible ideas is a way to find out
paths which will turn out to give the expected solution with, in many cases, surprising
outcomes. This paper aims at being a contribution to this large exchange. I hope that the
already few applications explored so far will be followed by many ones. I hope that it
will encourage people to venture along the tracks opened in this paper and, it would
be the best, to go further towards new avenues.

% practical problems are difficult
%  faire flèche de tout bois, à l'instar du système immunitaire qui bombarde les intrus
%  de tout un tas de substances jusqu'à ce que le bon anticoprs soit trouvé

% >>>>>>>>>>>>>>>>>>>>>>>>>>> Example of including a figure <<<<<<<<<<<<<<<<<<<<<<

%\begin{figure*}
%\centerline{
%\mbox{\includegraphics[width=2.00in]{engineeringletters.eps}}
%\mbox{\includegraphics[width=2.00in]{engineeringletters.eps}} }
%\caption{Replace text here with your desired caption.}
%\label{overView}
%\end{figure*}

% >>>>>>>>> Add a subsection if you want <<<<<<<<<<<<<<<<<<<<<<<<<<<<<<<<


\begin{thebibliography}{99}

% >>>>>>>>> Book examples <<<<<<<<<

\bibitem{ibkmIHM}
K. Chelghoum, M. Margenstern, B. Martin, I. Pecci,
Palette hyperbolique : un outil pour interagir avec des ensembles de
donn\'ees,
{\bf IHM'2004}, Namur, (2004). ({\it Hyperbolic chooser: a tool to interact with
data sets}, in French).

\bibitem{mmJUCSii}
Margenstern~M.,
New tools for cellular automata in the
hyperbolic plane, {\it Journal of Universal Computer Science}, vol {\bf 6},
issue 12, 1226--1252, (2000).

\bibitem{immwSCI}
Ch. Iwamoto, M. Margenstern, K. Morita, Th. Worsch,
Polynomial Time Cellular Automata in the Hyperbolic Plane Accept
Exactly the PSPACE Languages,
{\bf SCI'2002}, (2002).

\bibitem{mmASTC}
Margenstern~M.,
Implementing Cellular Automata on the Triangular Grids of the Hyperbolic
Plane for New Simulation Tools,
{\bf ASTC'2003}, (2003), Orlando, March, 29- April, 4.

\bibitem{cimmIEICE}
Iwamoto~Ch., Margenstern~M.,
Time and Space Complexity Classes of Hyperbolic Cellular Automata,
{\it IEICE Transactions on Information and Systems},
{\bf 387-D}(3), (2004), 700-707.

\bibitem{mmJCA}
Margenstern~M.,
A new way to implement cellular automata on the penta- and heptagrids,
{\it Journal of Cellular Automata} {\bf 1}(1), (2006), 1-24.

\bibitem{mmHypBio}
Margenstern~M.,
Can Hyperbolic Geometry Be of Help for P Systems?,
{\it Lecture Notes in Computer Science}, {\bf 2933}, (2004), 240-249.

\bibitem{mbuynACRI}
M. Margenstern, B. Martin, H. Umeo, S. Yamano, K. Nishioka,
A Proposal for a Japanese Keyboard on Cellular Phones,
{\it Lecture Notes in Computer Science}, {\bf 5191}, (2008), 299-306. 


\bibitem{mmJCAb}
Margenstern~M.,
On the communication between cells of a cellular automaton on the penta-
and heptagrids of the hyperbolic plane,
{\it Journal of Cellular Automata}
{\bf 1}(3), (2006), 213-232.


\bibitem{mmbook1}
Margenstern~M., 
{\it Cellular Automata in Hyperbolic Spaces, vol.~$1$: Theory},
Old City Publishing, Philadelphia, (2007), 422p.

\bibitem{mmJCAc}
Margenstern~M.,
A uniform and intrinsic proof that there are universal cellular
automata in hyperbolic spaces,
{\it Journal of Cellular Automata}, {\bf 3}(2), (2008), 157-180.

\bibitem{mmbook2}
Margenstern~M., 
{\it Cellular Automata in Hyperbolic Spaces, vol.~$2$: Theory},
Old City Publishing, Philadelphia, (2008), 360p.

\bibitem{mmRodos}
Margenstern~M.,
Universal Cellular Automata in Hyperbolic Spaces,
{\bf 13$^{\rm th}$ WSEAS International Conference on Computers}, 
Rodos, (2009), July 23-25, 83-89, ISBN 978-960-474-099-4.

\bibitem{bmKeyboard}
Martin~B., 
VirHKey: a VIRtual Hyperbolic KEYboard with gesture interaction and visual feedback 
for mobile devices. {\bf Mobile HCI}, (2005), 99-106.

\bibitem{mmkm99}
Margenstern~M., Morita~K.,
A Polynomial Solution for 3-SAT in the Space of Cellular
Automata in the Hyperbolic Plane, {\it Journal of Universal Computations and
Systems}, {\bf 5}, (1999), 563-573.

\bibitem{mmjsPPL}
Margenstern~M., Song~Y.,
A new universal cellular automaton on the pentagrid,
{\it Parallel Processing Letters}, {\bf 19}(2), (2009), 227-246.

\bibitem{mmjsENTCS}
Margenstern~M., Song Y.,
A universal cellular automaton on the ternary heptagrid,
{\it Electronic Notes in Theoretical Computer Science},
{\bf 223}, (2008), 167-185.

\bibitem{stewart}
Stewart~I., A Subway Named Turing, Mathematical Recreations in {\it Scientific
American}, (1994), 90-92.

% >>>>>>>>> Conference Proceedings Example <<<<<<<<<

% >>>>>>>>> Journal Example <<<<<<<<<<<<<<<<<<<<<<<<

\end{thebibliography}
\end{document}